# TWO-DIMENSIONAL NONLINEAR REGIME IN THE PLASMA WAKEFIELD ACCELERATOR[*]

A. G. Khachatryan[#], S. S. Elbakian, Yerevan Physics Institute, Yerevan, Armenia

*Abstract*

The effect of nonlinearity on plasma wake wave excited by a relativistic cylindrical charged bunch is investigated. It is shown that owing to the nonlinearity the amplitude of wake wave gets modulated in the longitudinal direction. The nonlinear wavelength in the excited field changes in the transverse direction with the result that the phase front is distored and a turbulence developed. The nonlinear phase front distortion may be compensated by radial change of unperturbed plasma density.

## 1 INTRODUCTION

The plasma-based accelerator concepts are presently actively developed both theoretically and experimentally. This is due to the ability of plasma to support large acceleration rates that will reach tens of GeV/m, far in excess of those attained in conventional accelerators. Charged bunches can be accelerated and focused by the field of relativistic plasma waves that are excited by relativistic charged bunches (Plasma Wakefield Accelerator (PWFA)). Both the linear wakefield and 1D nonlinear theories have been studied in sufficient detail (see e. g. overview in Ref. [1]). The allowance for finite transverse sizes of the drivers (that is more realistic case) and, accordingly, the transverse motion of plasma electrons complicate the treatment of the problem in the nonlinear regime. In the general case the analytical solution of this regime seems impossible and the use of numerical methods are usually required. Here we study the effect of nonlinearity on two-dimensional plasma wake waves as well as discuss the cause of radial steepening of the field shown in Ref. [2].

## 2 BASIC EQUATIONS

We shall consider nonlinear steady fields excited in plasma by a rigid cylindrical electron bunch in the framework of cold hydrodynamics with immobile ions. Let the bunch travel in $Z$ direction at the velocity $v_0$ close to that of light, and the distribution of charge in the bunch do not depend on the azimuthal angle $\theta$. Equations for non-zero components of plasma electrons momentum and electromagnetic field that describe the nonlinear wake-fields are (see also Refs. [2,3]):

[*]Work supported by the International Science and Technology Center
[#]Email: khachatr@moon.yerphi.am

$$\beta \partial P_z / \partial z - \partial \gamma_e / \partial z - \beta^2 E_z = 0, \quad (1)$$

$$\beta \partial P_r / \partial z - \partial \gamma_e / \partial r - \beta^2 E_r = 0, \quad (2)$$

$$-\partial H_\theta / \partial z + \beta \partial E_r / \partial z + \beta_r N_e = 0, \quad (3)$$

$$\nabla_\perp H_\theta + \beta \partial E_z / \partial z + \beta_z N_e + \beta \alpha = 0, \quad (4)$$

$$\beta \partial H_\theta / \partial z - \partial E_r / \partial z + \partial E_z / \partial r = 0, \quad (5)$$

$$N_e = N_p(r) - \alpha - \nabla_\perp E_r - \partial E_z / \partial z. \quad (6)$$

As usual, Eqs. (1) and (2) were derived taking into account that the curl of the generalized momentum is zero, $\beta^2 \mathbf{H} - rot\mathbf{P} = 0$, or in our case

$$\beta^2 H_\theta + \partial P_z / \partial r - \partial P_r / \partial z = 0. \quad (7)$$

Also we allow for radial variations of unperturbed plasma density. In Eqs. (1)–(6) $\gamma_e = (1+P_z^2+P_r^2)^{1/2}$ is a relativistic factor, $\beta_{z,r} = P_{z,r}/\gamma_e$ and $N_e = n_e/n_p(0)$ are respectively dimensionless components of velocity and density of plasma electrons, $n_p(r)$ is the unperturbed density of plasma electrons, $N_p = n_p(r)/n_p(0)$, $\alpha = n_b(z,r)/n_p(0)$, $n_b$ is the concentration of bunch electrons, $\beta = v_0/c$, $\nabla_\perp = \partial/\partial r + 1/r$. The space variables are normalized to $\lambda_p(r=0)/2\pi = 1/k_p(r=0)$, $z = k_p(r=0)(Z - v_0 t)$, where $\lambda_p$ and $k_p$ are the linear wavelength and wavenumber. The momenta and velocities are normalized respectively to $m_e c$ and the velocity of light and the strengths of electric and magnetic fields - to the nonrelativistic wave-breaking field at the axis $E_{WB}(r=0) = m_e \omega_{pe}(r=0) v_0/e$ ($E_{WB}$[V/cm]$\approx 0.96 \times n_p^{1/2}$[cm$^{-3}$]), $\omega_{pe} = (4\pi n_p e^2/m_e)^{1/2}$ is the electron plasma frequency, $m_e$ and $e$ are the rest mass and absolute value of electron charge. The field of forces acting on relativistic electrons in the excited field is $\mathbf{F}(-eE_z, e(H_\theta - E_r), 0)$.

## 3 THE CASE OF WIDE BUNCH

Consider the case of wide bunch ($k_p r_b \gg 1$, where $r_b$ is the bunch radius) when the transverse components of an exciting field are small and the longitudinal components close to the bunch axis are approximately equal to those predicted by one-dimensional nonlinear theory. Here one can apply the perturbation method taking the 1D nonlinear mode as the ground state. So, for wide bunches in uniform plasma [$N_p(r)=1$] we shall seek the solution of Eqs. (1)–(6) in the vicinity of bunch axis in the form: $P_z \approx P_0 + \rho_z r^2$, $P_r \approx \rho_r r$, $E_z \approx E_0 + l_z r^2$, $E_r \approx l_r r$, $H_\theta \approx hr$, $\alpha \approx \alpha_0 - \delta r^2$, where $P_0(z)$ and $E_0(z)$ are the values of longitudinal momentum and the strength of electric field at the axis, $\rho_{z,r}(z)$, $l_{z,r}(z)$, $h(z)$, $\delta(z) \ll 1$. In the zero approximation in this values one can obtain the equations for $P_0$ and $E_0$ that describe the 1D nonlinear regime that is studied sufficiently well [4]. In the first approximation in the small values we have:

$$d^2\rho_z/dz^2 - Ad\rho_z/dz + B\rho_z = \beta N_0\delta, \quad (8)$$

where $A=(d\beta_0/dz)2N_0/\beta$, $B=(N_0/\beta)(\beta N_0^2/\gamma_0^3 - d^2\beta_0/dz^2)$, $\beta_0$, $N_0$ and $\gamma_0$ are respectively dimensionless velocity, density and the relativistic factor of plasma electrons in 1D nonlinear regime. The remaining quantities can be expressed through $\rho_z$. In Fig. 1 we show the focusing gradient $f_r/r=(H_\theta-E_r)/r=-2\rho_z/N_0$, that is excited by $d=4.7$ long wide bunch with density $\alpha_0=0.2$.

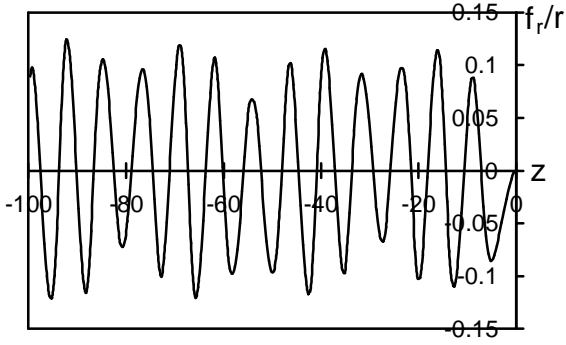

Figure 1: The gradient of dimensionless focusing field as a function of longitudinal coordinate. The bunch is $d=4.7$ long, and $\alpha_0=0.2$, $\delta=0.2\alpha_0$, $\gamma=(1-\beta^2)^{-1/2}=10$.

It was shown numerically based on Eq. (8) that although the amplitude of excited transverse components is small, their wavelength is nearly equal to that of one-dimensional nonlinear wave. The amplitude of oscillations periodically changes with $z$ (see Fig. 1). The modulation depth grows and the modulation period decreases as the amplitude of nonlinear longitudinal oscillations increases. These effects take place also for other components of the field.

## 4 WAKEFIELD IN UNIFORM PLASMA

Eqs. (1)–(6) were solved numerically choosing the Gaussian profile of the bunch both in longitudinal and transverse directions:

$$\alpha = \alpha_0 \exp[-(z-z_0)^2/\sigma_z^2]\exp(-r^2/\sigma_r^2). \quad (9)$$

In case of small amplitudes of the excited wake wave (when $\alpha_0 \ll 1$) the numerical solutions well agreed with linear theory predictions. Shown in Fig. 2 is the nonlinear 2D plasma wake wave excited in uniform plasma by the relativistic electron bunch with parameters $\alpha_0=0.4$, $\sigma_z=2$, $\sigma_r=5$ (for example, in this case $n_{b0}=4\times10^{13}\text{cm}^{-3}$ and the characteristic longitudinal and transverse sizes of the bunch $\sigma_z/k_p$ respectively are 1.06mm and 2.65mm when $n_p=10^{14}\text{cm}^{-3}$). The main difference here from the linear case is the change of shape and length of the wave with the radial coordinate $r$, as well as the change of amplitude in the longitudinal direction. Note also an enlargement of the range of focusing forces ($f_r<0$) in the nonlinear wake wave.

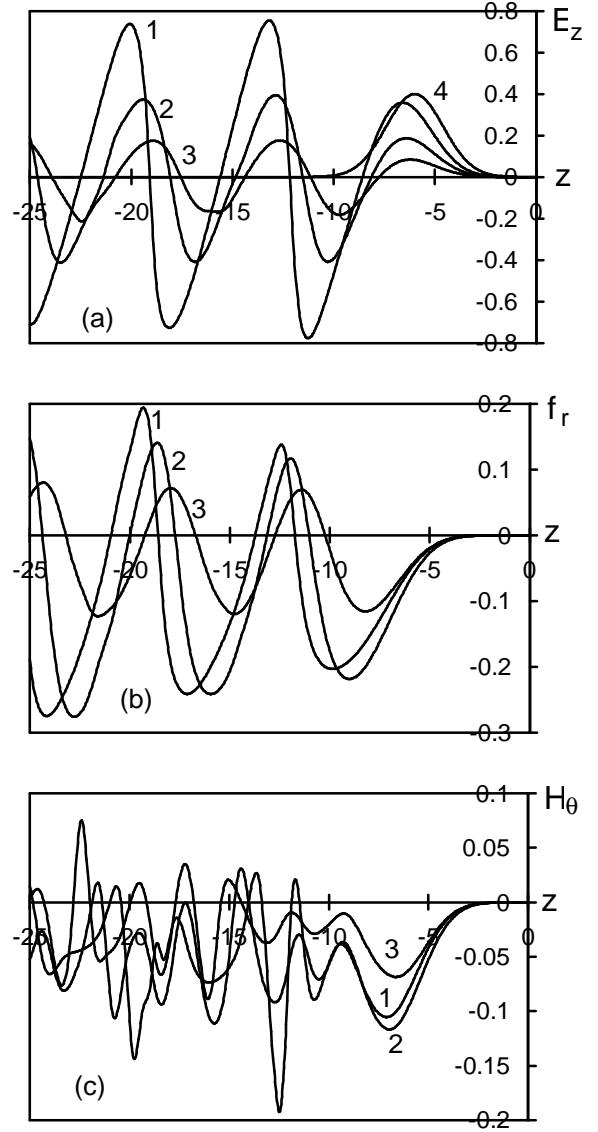

Figure 2: The two-dimensional nonlinear wake wave in uniform plasma. The parameters of the bunch are: $\alpha_0=0.4$, $\sigma_z=2$, $\sigma_r=5$, $\gamma=10$. (a). The longitudinal electric field. 1–$r=0$; 2–$r=4$; 3–$r=6$; 4–the density of bunch at the axis $\alpha(z, r=0)$. The focusing field $f_r=H_\theta-E_r$ (b) and magnetic field strength (c). 1–$r=2$; 2–$r=4$; 3–$r=6$.

It is easy to see that due to the dependence of the wavelength on $r$, the field in the radial direction grows more chaotic as the distance from the bunch increases. In fact, the oscillations of plasma for different $r$ are "started" in the wake wave with nearly equal phases but different wavelengths. As $|z|$ increases, the change of phase in the transverse direction (for fixed $z$) becomes more and more marked. This leads to a curving of the phase front and to "oscillations" in the transverse direction. Curve 1 in Fig. 3 shows the radial behavior of the accelerating field $E_z$ in the nonlinear wake wave. Qualitatively, the radial dependence of the field differs from that of the linear case by the change of sign and "steepening" of fields along $r$ (see also Ref. [2]). One can determine the longitudinal space parameter characterizing the wave front curving as follows:

$$\xi = \lambda_p / [2(1 - \lambda_p / \Lambda(0))], \quad (10)$$

where $\Lambda(r)$ is the nonlinear wavelength. At the distance $|\Delta z| \approx \xi$ from the bunch the oscillation phase at the axis ($r=0$) is opposite to that on the periphery ($r \geq \sigma_r$) where the wave is nearly linear.

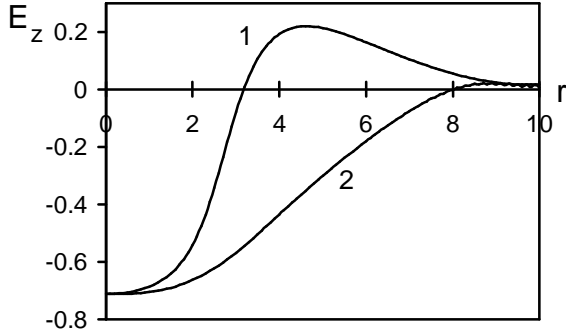

Figure 3: The radial behavior of accelerating electric field strength $E_z$. 1–$E_z(z=-25, r)$ in the nonlinear wake wave excited in uniform plasma for the case given in Fig.2 ($|\Delta z| \approx \xi$ [see Eq. (10)]); 2–$E_z(z=-25, r)$ in nonuniform plasma for the case given in Fig. 4.

The magnetic field strength in the nonlinear two-dimensional wake wave as shown in Fig. 2(c) is different from zero (note that in the linear case $H_\theta=0$ in the wake, where the wave is potential) due to the fact that the velocities of plasma electrons in the wave are not any more small in comparison with the bunch velocity. The magnitude of higher frequency oscillations (as compared to the plasma frequency) performed by the magnetic field along $z$ grows in proportion to the nonlinearity. Such a behavior of magnetic field is a purely nonlinear effect. The nonlinearity of the wave implies a rise of higher harmonics in $P_z$ and $P_r$. According to (7), the rise of magnetic field is due to these harmonics and this accounts for frequent oscillations seen in Fig. 2(c). On the other hand, according to (7), the non-zero magnetic field in the wake means that the motion of plasma electrons in the nonlinear wave is turbulent ($rot\mathbf{P} \neq 0$). The degree of turbulence (the measure of which is $H_\theta$) grows as the nonlinearity.

From the viewpoint of acceleration and focusing of charged bunches in the wave, the curvature of the nonlinear wave front is undesired as the quality (emittance, monochromaticity) of the driven bunch worsens. Below we show that in radially-nonuniform plasma one can avoid the curvature of the wave front.

## 5 RADIALLY-NONUNIFORM PLASMA

Thus, in two-dimensional nonlinear regime the nonlinear wavelength changes with $r$ due to nonlinear increase of the wavelength with wave amplitude. On the other hand, the linear wavelength $\lambda_p \sim n_p^{-1/2}$ decreases with density of plasma. Let us assume that the nonlinear wavelength of the two-dimensional wake wave in the uniform plasma $\Lambda(r)$ is known. Then, one can roughly compensate for the radial variation of the nonlinear wavelength by changing the unperturbed density of plasma in the radial direction according to the relation

$$\Lambda(0)/\Lambda(r) = \lambda_p(r)/\lambda_p(0) = [n_p(0)/n_p(r)]^{1/2}. \quad (11)$$

If we put the function $\Lambda(r)$ to be Gaussian (according to numerical data for profiles (9), this is approximately the case at least for $r<\sigma_r$), then one can take the transverse profile of the unperturbed plasma density to be also Gaussian: $n_p(r) = n_{p0} \exp(-r^2/\sigma_p^2)$. It follows from Eq. (11) that in this case $\sigma_p = r/[\ln(\Lambda(0)/\Lambda(r))]^{1/2}$. For example, the numerical data for $\Lambda(r)$ in the nonlinear wave shown in Fig. 2 give $\sigma_p \approx 11$. Fig. 4 illustrates the validity of this assertion (see also Fig. 3, curve 2).

Numerical solutions obtained for the case of nonlinear wake wave excitation by a short laser pulse (Laser Wakefield Accelerator, see e. g. Ref. [1]) show that the results given in this work do not change qualitatively.

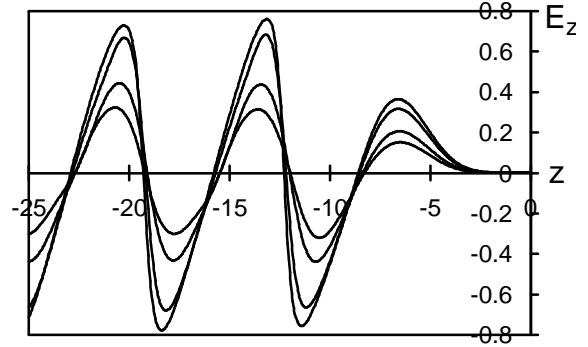

Figure 4: Accelerating electric field of two-dimensional nonlinear wake wave in nonuniform plasma with $\sigma_p=11$ for $r=0$, 2, 4 and 5 in the order of magnitude reduction. The bunch parameters are the same as those in Fig. 2.

## 6 REFERENCES

[1] E. Esarey, P. Sprangle, J. Krall, and A. Ting, IEEE Trans. Plasma Sci. **24**, 252 (1996).

[2] B. N. Breizman, T. Tajima, D. L. Fisher, and P. Z. Chebotaev, In: *Research Trends in Physics: Coherent Radiation and Particle Acceleration, edited by A. Prokhorov* (American Institute of Physics, New York, 1992), pp. 263-287.

[3] K. V. Lotov, Phys. Plasmas **5**, 785 (1998).

[4] A. Ts. Amatuni, E. V. Sekhpossian, and S. S. Elbakian, Fiz. Plasmy **12**, 1145 (1986); J. B. Rosenzweig, Phys. Rev. Lett. **58**, 555 (1987); A. G. Khachatryan, Phys. Plasmas **4**, 4136 (1997).